\documentclass[12pt]{iopart}

\usepackage{graphicx}
\usepackage{color}
\usepackage{iopams}

\begin{document}

\title{Nuclear matter response function with a central plus tensor Landau interaction}

\author{A. Pastore}
\address{Institut d'Astronomie et d'Astrophysique, CP 226, Universit\'e Libre de Bruxelles, B-1050 Bruxelles, Belgium}

\author{D. Davesne}
\address{Universit{\'e} de Lyon, F-69622 Lyon, France, \\
Universit\'e Lyon 1, Villeurbanne;  CNRS/IN2P3, UMR5822, Institut de Physique Nucl\'eaire de Lyon}

\author{J. Navarro}
\address{IFIC (CSIC-Universidad de Valencia), Apartado Postal 22085, E-46.071-Valencia, Spain}

\date{\today}

\begin{abstract}
We present a method to obtain response functions in the random phase approximation (RPA) based on a residual interaction described in terms of Landau parameters with central plus tensor contributions. The response functions keep the explicit momentum dependence of the RPA, in contrast with the traditional Landau approximation. Results for symmetric nuclear matter and pure neutron matter are presented using Landau parameters derived from finite-range interactions, both phenomenological and microscopic. We study the convergence of response functions as the number of Landau parameters is increased. 
\end{abstract}
\pacs{21.30.-x, 21.60.Jz, 21.65.+f}

\maketitle

\section{Introduction}

Infinite nuclear matter is a very useful and broadly used concept, which can provide insights 
about the inner part of atomic nuclei, and model some regions of compact stars. It has long been an important benchmark for phenomenological and microscopic nuclear interactions and for many-body methods as well. 
The response of nuclear matter to different probes is the key quantity to understand the 
excitation of collective states or the onset of instabilities to density fluctuations. 
As an example of astrophysical interest we mention the study of neutrino transport properties in dense matter, which requires a detailed knowledge of the nuclear response.

The Landau's theory \cite{mig67,bay91,noz99} encompasses the basic properties of Fermi liquids, and is a simple method largely used to calculate response functions of nuclear matter. The excitations of a strongly interacting normal Fermi system are described in terms of weakly interacting quasiparticles --or particle-hole (ph) excitations-- which are long-lived only near the Fermi surface. The responses are obtained by solving the linearized Boltzmann transport equation in the long-wavelength limit. Denoting the transferred energy and momentum by $\omega$ and $q$,  respectively, the so-called Landau limit consists in taking $q \to 0$, but keeping the quotient $\omega/q$ fixed. These response functions depend only on the dimensionless variable $\omega /(v_F q)$, where  $v_F$ is the Fermi velocity. 
(Natural units are used along this paper). The quasiparticle interaction is solely characterized by a set of  Landau parameters, which can be obtained from phenomenological or realistic interactions. 

Another method to obtain the response function is provided by the random-phase approximation (RPA), which is the small amplitude limit of a time-dependent mean-field approach \cite{fet71,bla86,lip03}. The RPA response is a function of the two variables $\omega$ and $q$, and not only of the quotient $\omega/q$. 
In the theory of Fermi liquids, the ph interaction is given by the second functional derivative of the total energy with respect to densities taken at the Hartree-Fock (HF) solution. Exchange terms are directly included in the ph interaction by this procedure~\cite{gar92}. If the total energy  is obtained from an energy density functional (EDF), the RPA response is self-consistent in the sense that the mean field and the ph interaction are derived from the same energy functional. Detailed calculations have been mostly done based on phenomenological interactions \cite{eng99}. 

In this paper we consider a third method to obtain the response, which is an intermediate possibility between the previous two. It consists in describing the ph interaction in terms of Landau parameters, including both central and tensor components, and calculating the RPA response function with no further approximations. Actually we shall apply and extend the method previously developed to obtain the RPA response function \cite{gar92,mar06,dav09,nav13}. We shall explore the convergence of the response in terms of the number of Landau parameters. Our aim is to provide a general tool to obtain the response function in a simple way, which can be of interest for astrophysical applications. 

The three mentioned approaches have been used in different physical situations, considering a small number of Landau terms, but with no tensor contributions. Consider for instance the neutrino mean-free path in dense matter. It has been calculated in the framework of the Landau approximation using the first two central Landau parameters
\cite{iwa82,ben13}, and also using the RPA response function either based on Skyrme interactions \cite{nav99}, or 
 a residual interaction described with one or two Landau parameters \cite{red99,she03,mar03}. The Landau parameters have been calculated in a large variety of ways: from a G-matrix using the Reid soft-core interaction, in the formalism of the correlated basis function (CBF) using the Argonne AV18 interaction, from phenomenological interactions of Skyrme or Gogny type, or extracted from Brueckner-Hartree-Fock approach using the AV18 interaction.  
New calculations of  Landau parameters including both central and non-central components  are now available from microscopic calculations based on high-precision two- and three-body forces \cite{ben13,ols04,sch04,hol13,pet09}. 
They can provide useful guides for the construction of phenomenological interactions of finite range, as Gogny \cite{gogny} or M3Y \cite{nak03}, and zero range as Skyrme~\cite{cha97}. In this case, the Landau parameters are used to impose stability constraints \cite{cao} to prevent unphysical phase transitions not predicted by \emph{ab-initio} calculations~\cite{nav13,mar02}. 

Guided by the recent interest on the possible extensions of Skyrme interactions in powers of momentum operators~\cite{car08,dav13} and the inclusion of zero-range three-body forces~\cite{sad13}, we have investigated the RPA  response function in infinite systems using the Landau ph interaction. The article is organized as follows: in Sec ~\ref{formalism} we present the general formalism to obtain the response function using central and tensor Landau parameters. In Sec. \ref{results}, we present our numerical results in the different spin and isospin channels, for several choices of parameters. The summary and conclusions are given in Sec. \ref{conclusion}.

\section{Formalism}
\label{formalism}

\subsection{Particle-hole interaction}\label{res:int}
 
Due to momentum conservation, a general two-body interaction in momentum representation depends at most on three momenta. For the particle-hole case we choose them to be the initial (final) momentum ${\mathbf k}_1$ (${\mathbf k}_2$) of the hole and the external momentum transfer ${\mathbf q}$.
In the Landau-Migdal approximation \cite{mig67} it is assumed that the low-energy excitations of the system are described by putting the interacting particles and holes on the Fermi surface, that is $k_1 = k_F$, $k_2=k_F$, $q=0$. The only variable is thus the relative angle $({\mathbf {\hat k}_1} \cdot {\mathbf {\hat k}_2})$ between the initial and final momenta. 

The \emph{p-h} interaction is in practice a contact interaction, which is expanded in Legendre polynomials with argument $({\mathbf {\hat k}_1} \cdot {\mathbf {\hat k}_2})$. It includes spin and isospin degrees of freedom, and the general form for symmetric nuclear matter (SNM) adopted here reads:
\begin{eqnarray}
\label{landau-Vph}
 V_{ph}  &=&  \sum_{\ell} \,
\bigg\{ f_{\ell} + f_{\ell}' \, (\btau_1 \cdot \btau_2 ) 
+ \left[ g_{\ell}+ g_{\ell}' \, (\btau_1 \cdot \btau_2) 
\right] (\bsigma_1 \cdot \bsigma_2)  \\
&& \quad  + \left[ h_{\ell} + h_{\ell}' \, (\btau_1 \cdot \btau_2 ) \right]  \frac{{\mathbf k}_{12}^{2}}{k_{F}^{2}} \, S_{12}
\bigg\} \, P_{\ell} ( {\mathbf {\hat k}_1} \cdot {\mathbf {\hat k}_2}  ) \nonumber
\end{eqnarray}
where $f_{\ell}, f'_{\ell}, \dots$ are the Landau parameters, ${\mathbf k}_{12} = {\mathbf k}_1 - {\mathbf k}_2$, and $S_{12}= 3( \hat{\mathbf{k}}_{12} \cdot \bsigma_1)( \hat{\mathbf{k}}_{12} \cdot \bsigma_2 )- ( \bsigma_1 \cdot \bsigma_2) $ is the tensor operator. We do not include other non-central components, as the center-of-mass tensor and cross-vector interactions, which have been considered by other authors (see for instance \cite{ols04,sch04}).
 
The excitations are characterized by the spin-isospin quantum numbers $(S, M, I, Q)$, where $M$ and $Q$ refer to the projections of the spin $S$ and isospin $I$, respectively. As long as we are not interested in charge-exchange processes, the isospin projection index $Q$ is irrelevant and will be ignored. 
For pure neutron matter (PNM) there are only two quantum numbers, $(S,M)$ but the interaction (\ref{landau-Vph}) can be adapted in a very simple way to this case  by dropping the coefficients $f'_{\ell}, g'_{\ell}, h'_{\ell}$ and using the notation $f^{(n)}_{\ell}, g^{(n)}_{\ell}, h^{(n)}_{\ell}$ for the remaining ones.
In the following, all the expressions for SNM and PNM will be cast in a single formal expressions by using the symbol $(\alpha)$ to indicate the relevant spin-isospin quantum numbers. For instance, the Landau parameters $f_{\ell}, f'_{\ell}, g_{\ell}, g'_{\ell}$ will be written as $f_{\ell}^{(\alpha)}$, with $(\alpha) = (0,0), (0,1), (1,0), (1,1)$, respectively. These parameters are independent of the spin projection $M$. 
The dimensionless parameters $F_{\ell}^{(\alpha)}$ are obtained by multiplying the previous
ones with the density of states per energy at the Fermi surface $N(0) = n_d
k_F m^* /(2 \pi^2)$, where $n_d$ is the spin-isospin degeneracy factor, i.e., 4 for SNM and 2 for PNM. 

The tensor part in (\ref{landau-Vph}) has been written according to~\cite{dab76,bac79}, which is the conventional definition. Other authors \cite{ols04,sch04,ben13,hol13} have defined it without the factor ${\bf k}_{12}^2/k_F^2$. Although the physical information is the same, the Landau parameters are different in that case. This particular choice is motivated by a faster convergence \cite{ols04}, in the sense that the absolute value of parameters $h_{\ell}, h'_{\ell}$ decreases as $\ell$ increases. However the conventional definition is more adapted to our method to calculate the response function so that we will keep it throughout this paper. We refer to~\cite{ols04} for a more detailed discussion.

\subsection{RPA response function}\label{res:function}
We employ the method and notations presented in Refs. \cite{gar92,mar06,dav09,nav13} to obtain the RPA response function. It requires the matrix elements of the ph interaction (\ref{landau-Vph}) between spin-isospin states $V_{ph}^{(\alpha,\alpha')}=\langle \alpha| V_{ph}|\alpha'\rangle$. Notice that they are diagonal in the indices $S$ and $I$. The non-diagonal terms are related to the spin projection index $M$, since the tensor term mixes its different values.  We can therefore write them as:
\begin{eqnarray}
\label{Vph-ME}
\fl V_{ph}^{(\alpha,\alpha')}/n_d &=& \delta(M,M') \sum_{\ell}   f^{(\alpha)}_{\ell} 
P_{\ell}( {\mathbf {\hat k}_1} \cdot {\mathbf {\hat k}_2}  ) 
 + \delta(S,1) \sum_{\ell} h^{(\alpha)}_{\ell}  
P_{\ell}( {\mathbf {\hat k}_1} \cdot {\mathbf {\hat k}_2}  ) 
S_T^{(M,M')}( {\mathbf {\hat k}_1} , {\mathbf {\hat k}_2}  ) 
\end{eqnarray}
where
\begin{eqnarray}
S_T^{(M,M')}( {\mathbf {\hat k}_1} , {\mathbf {\hat k}_2}  ) =  
3  (-)^M \left( k_{12}\right)^{(1)}_{-M} \left( k_{12}\right)^{(1)}_{M'}
- \delta(M,M') 2 \, \left[ 1- ( {\mathbf {\hat k}_1} \cdot {\mathbf {\hat k}_2}  ) \right] 
\end{eqnarray}
and following the notation of \cite{dav09} we have defined
\begin{eqnarray}
\left( k_{12}\right)^{(1)}_{M} = \sqrt{\frac{4 \pi}{3}} \left( Y_{1,M}({\mathbf {\hat k}_1}) -
Y_{1,M}({\mathbf {\hat k}_2}) \right).
\end{eqnarray}
The product $\delta(S,S') \delta(I,I')$ is implicitly assumed in the r.h.s. of  (\ref{Vph-ME}). 

These matrix elements are plugged into the Bethe-Salpeter equation for the retarded ph propagator
\begin{eqnarray}
\label{bsalpeter}
\fl G^{(\alpha)}_{RPA}({\bf q},\omega,{\bf k}_1)
&=& G_{HF}({\bf q},\omega,{\bf k}_1) +\sum_{(\alpha')} G_{HF}({\bf q},\omega,{\bf k}_1) \int d{\mathbf{k}}_{2}  V_{ph}^{(\alpha, \alpha')}  G^{(\alpha')}_{RPA}({\bf q},\omega,{\bf k}_2)
\end{eqnarray}
where
\begin{equation}
G_{HF}({\bf q},\omega,{\bf k}_1) = 
\frac{\theta(k_F-k_1) - \theta(k_F-|{\bf k}_1 + {\bf q}|)}{\omega+\varepsilon(k_1) -
\varepsilon(|{\bf k}_1 + {\bf q}|) + i \eta} \, ,
\end{equation}
is the Hartree-Fock (HF) ph propagator. Finally, the response function is obtained as 
\begin{equation}
\chi_{RPA}^{(\alpha)}({\bf q},\omega) = n_d \int \frac{d{\mathbf{k}}_{1}}{(2 \pi)^3} 
G^{(\alpha)}_{RPA}({\bf q},\omega,{\bf k}_1)
\end{equation}
To simplify the notation, momentum averages as the previous one will be indicated within brackets as $\chi_{RPA}^{(\alpha)}({\bf q},\omega) = n_d \langle G^{(\alpha)}_{RPA} \rangle$. 

To obtain the response function one has to take the momentum average of (\ref{bsalpeter}). The second term in the rhs is thus transformed into a double average on momenta $\mathbf{k_1}$ and $\mathbf{k_2}$.
To understand how we can treat such an expression, let us consider the dependence of angles ${\mathbf {\hat k}_1}$ and ${\mathbf {\hat k}_2}$ entering the ph interaction (\ref{Vph-ME}). This dependence comes from the Legendre polynomials which are linear combinations of
$( {\mathbf {\hat k}_1} \cdot {\mathbf {\hat k}_2}  )^n$, which in turn are combinations of the products
$Y^*_{1,\mu_1}( {\mathbf {\hat k}_1}) \dots Y^*_{1,\mu_n}( {\mathbf {\hat k}_1})
Y_{1,\mu_1}( {\mathbf {\hat k}_2}) \dots Y_{1,\mu_n}( {\mathbf {\hat k}_2})$. 
Due to the tensor part of the interaction, these combinations may appear multiplied by 
one of the following terms 
$$Y^*_{1,M}( {\mathbf {\hat k}_1}) Y_{1,M'}( {\mathbf {\hat k}_1}), \,  
Y^*_{1,M}( {\mathbf {\hat k}_1}) Y_{1,M'}( {\mathbf {\hat k}_2}), \, 
Y^*_{1,M}( {\mathbf {\hat k}_2}) Y_{1,M'}( {\mathbf {\hat k}_1}), \,  
Y^*_{1,M}( {\mathbf {\hat k}_2}) Y_{1,M'}( {\mathbf {\hat k}_2}).$$
Therefore, the momentum average we are interested in, $\langle G^{(\alpha)}_{RPA} \rangle$, is coupled to other averages containing a certain number of spherical harmonics $\langle Y_{1,m_1}( {\mathbf {\hat k}_2}) \dots Y_{1,m_i}( {\mathbf {\hat k}_2})
G^{(\alpha)}_{RPA} \rangle$. The idea consists in multiplying the Bethe-Salpeter equation by these products of spherical harmonics and integrating over the momenta to get a new equation for each of these averages, until one ends up with a closed system of coupled linear equations for these unknown functions. The coefficients depend on the Landau parameters and momentum averages of the HF propagator.  

Let us illustrate the method in the first non-trivial case of a ph interaction characterized by the first two Landau central parameters. To be more specific consider the $S=0, T=0$ channel of SNM.
Integrating the Bethe-Salpeter equation we get:
\begin{eqnarray}
\fl \langle G_{RPA}^{(0,0)} \rangle = \langle G_{HF} \rangle 
+ n_d f_0 \langle G_{HF} \rangle \langle G_{RPA}^{(0,0)} \rangle 
+ n_d f_1 \frac{4 \pi}{3} \sum_{\mu}  \langle Y^*_{1,\mu} G_{HF} \rangle \langle Y_{1,\mu} G_{RPA}^{(0,0)} \rangle .
\end{eqnarray}
Without loss of generality the vector $\mathbf{q}$ can be chosen along the $z$-axis. Since the HF ph propagator $G_{HF}$ does not depend on the azimuthal angle $\phi$, the momentum average $\langle Y^*_{1,\mu} G_{HF} \rangle$ vanishes unless $\mu=0$. The previous equation can thus be written as:
\begin{eqnarray}
\langle G_{RPA}^{(0,0)}  \rangle = \alpha_0 + n_d f_0 \alpha_0 \langle G_{RPA}^{(0,0)} \rangle + n_d f_1  \alpha_1 \langle \cos \theta \; G_{RPA} ^{(0,0)} \rangle \, ,
\label{f01-1}
\end{eqnarray}
where we have defined:
\begin{eqnarray}
\alpha_i(q, \omega) = \langle \cos^i \theta \, G_{HF} \rangle .
\end{eqnarray}
The function $\langle G_{RPA}^{(0,0)} \rangle$ we are looking for is coupled to $\langle \cos \theta \; G_{RPA}^{(0,0)} \rangle$. A second equation for this quantity is then obtained by multiplying the Bethe-Salpeter equation with $\cos \theta_1$ and integrating over $\mathbf{k}_1$: 
\begin{eqnarray}
\langle \cos \theta \, G_{RPA} ^{(0,0)} \rangle = \alpha_1 + n_d f_0 \alpha_1 \langle G_{RPA}^{(0,0)}  \rangle 
+ n_d f_1  \alpha_2 \langle \cos \theta \, G_{RPA} ^{(0,0)}  \rangle .
\label{f01-2}
\end{eqnarray}
Equations (\ref{f01-1}) and (\ref{f01-2}) form a closed coupled system, from which one immediately obtains $\langle G_{RPA} ^{(0,0)} \rangle$. Finally, the RPA response reads:
\begin{eqnarray}
\label{RPA-01}
\frac{\chi_{HF} (q,\omega) }{\chi_{RPA} ^{(0,0)}(q,\omega) } = 1 - \left[ f_0 + f_1 \frac{\alpha_1^2/\alpha_0^2}{1-n_d f_1 (\alpha_0 \alpha_2 - \alpha_1^2)/\alpha_0} \right] \chi_{HF} (q,\omega) 
\end{eqnarray}
This expression is also valid for the other $(\alpha)$-channels, both in SNM and PNM, by simply using the appropriate Landau parameters.
Notice that $\chi_{HF} = n_d \alpha_0$. The functions $\alpha_i(q,\omega)$ carry on the mean-field information of the response function. They play a similar role than the functions $\beta_{i}(q,\omega)$ introduced in Ref.~\cite{gar92}, the difference is that the latter include a $k$-dependence in the momentum integral. Analytic expressions of the imaginary part of  $\alpha_{i=0,8}(q,\omega)$ functions are given in \ref{app:alpha} together with a typical example of their graphical representation.

This simple case illustrates our method for solving the Bethe-Salpeter equation.
It has been already employed in the case of a general  ph  interaction derived from a standard Skyrme effective interaction including tensor components \cite{dav09}  and then generalized in \cite{pas12a,pas12b} to the case of a second order Skyrme functional~\cite{per04}. The complete ph interaction has a complicated momentum dependence, but due to its zero-range nature  it is possible to obtain practical analytical expressions for the response function in that case. 
In general, the algebraic system of coupled equations can be simplified to get practical analytical expressions only in the short-frequency and long-wavelength limits, that is to calculate the 
static susceptibility, as was done in \cite{nav13} considering a ph interaction (\ref{Vph-ME}) with an infinite number of terms. Otherwise, to obtain the RPA response function with more than two Landau parameters it is preferable to numerically solve the algebraic system derived from the Bethe-Salpeter equation.

\subsection{Landau approximation for the response function}
The consistency of our method can be tested by taking the Landau limit $q \to 0$ and $\omega/q$ finite in 
the RPA response (\ref{RPA-01}). In that limit $\alpha_1 \to \nu \alpha_0$, and  $\alpha_2 \to \nu^2 \alpha_0 - N(0) / (3 n_d)$, where $\nu= \omega m^*/( q k_F)$. We obtain thus the familiar expression for the Landau response function:
\begin{eqnarray}
\label{LAN-01}
\frac{\chi_{HF}(\nu)}{\chi_{RPA}^{(0,0)}(\nu)} = 1 - \left[ f_0 + \frac{f_1 \nu^2}{1+ F_1/3} \right] \chi_{HF}(\nu) \, ,
\end{eqnarray}
 as it should.

In the Landau limit, the standard Skyrme interaction leads to a ph interaction described with two central and one tensor parameters in the spin channel. Since an analytical expression for the RPA response function with the full Skyrme interaction has been obtained \cite{dav09}, one may expect to get a relatively simpler form when using its approximated version. The systems for channels $(S=1, M=0)$ and $(S=1, M=1)$ contain four and five equations, respectively.  However, contrarily to the case with the complete Skyrme interaction, it is unpractical to write the analytical expressions for the response functions in terms of the Landau parameters. This apparent paradox is related to the fact that the $\beta_i$ functions used in the former case fulfill some symmetry relations which help a lot in simplifying the final expressions. In contrast, there are no such relations for the $\alpha_i$ functions, and the analytical expressions are long and cumbersome. Nevertheless we have exploited them to derive their Landau limit. 

For channel $(S=1, M=0)$: 
\begin{eqnarray}
\fl \frac{\chi_{HF}(\nu)}{\chi_{RPA}(\nu)} 
&=& (1+H_0)^2  \\
\fl &-& \left( g_0 - 2 h_0 (1-  3 \nu^2) - 3 H_0 h_0 (1- \nu^2)  
+ \frac{ \nu^2 (g_1 - 4 h_0) (1+X^{(0)}) }{1+X^{(0)} + 
\frac{1}{3} ( G_1- 4 H_0)} \right) \chi_{HF}(\nu) \nonumber
\label{LAN-S10}
\end{eqnarray}
with
\begin{equation}
 X^{(0)} = \frac{  h_0 H_0 \left( 12(1-\nu^2) \alpha_0 +N(0) \right)}
{1- \frac{1}{6} (g_1 -7 h_0) \left( 12(1-\nu^2) \alpha_0 +N(0) \right)} \, ,
\end{equation}
which can be further simplified for small values of $\nu$ as
\begin{equation}
 X^{(0)} = \frac{ 3 H_0^2 (\nu^2 -2/3)}
{1- \frac{1}{2} (G_1 -7 H_0) (\nu^2 -2/3)} \, .
\end{equation}
Analogously, for channel $(S=1, M= \pm 1)$:
\begin{eqnarray}
\fl \frac{\chi_{HF}(\nu)}{\chi_{RPA}(\nu)} 
&=& (1 - \frac{1}{2} H_0)^2 \\
\fl &-& \left( g_0 + h_0 (1- 3 \nu^2) - \frac{9}{4} H_0 h_0 ( \nu^2 -1)^2   
+ \frac{\nu^2 (g_1 + 2 h_0)(1 +  X^{(1)})}{1 +  X^{(1)} + 
\frac{1}{3} (G_1 + 2 H_0)} \right) \chi_{HF}(\nu) \nonumber 
\label{LAN-S11}
\end{eqnarray}
with
\begin{equation}
 X^{(1)} = \frac{ \frac{3}{2} H_0^2 (\nu^2 -2/3)}
{1- \frac{1}{2} (G_1 + 2 H_0) (\nu^2 -2/3)} \, .
\end{equation}

Similarly to Eq. (\ref{RPA-01}), these expressions are also valid for other channels, both in SNM and PNM, by simply replacing the appropriate Landau parameters.
To the best of our knowledge, these expressions have never been obtained by solving the linearized Boltzmann equation. 
Analytical formulae have been given in that way only for the  static susceptibility, that is the long-wavelength and short-frequency limits, for a system of electrons \cite{fuj87} and, independently, for PNM \cite{ols04} as well. 

\section{Results}\label{results}

In this section, we calculate response functions for different sets of Landau parameters extracted from finite-range interactions.  In the following we prefer to show the strength function, related to the response function by:
\begin{equation}
S_{RPA}^{(\alpha)}({\bf q},\omega) = - \frac{1}{\pi} {\rm Im} \chi_{RPA}^{(\alpha)}({\bf q},\omega) \,
\end{equation}
since all physical properties are actually embedded into it.  

\subsection{Choice of interactions}
\label{params}

In principle, any finite-range interaction gives non-vanishing Landau parameters for any value of $\ell$, so that Eq. (\ref{Vph-ME}) contains an infinite number of terms. In contrast, the only non-vanishing Landau parameters calculated with standard Skyrme interactions are $\ell=0, 1$ for the central ones, and  $\ell=0$ for the tensor one. For practical use, such an expansion has to be truncated to a maximum value, independent in principle for the central and the tensor parts. 
Our choice for the truncation is guided by the recent Skyrme N3LO pseudo-potential \cite{car08,rai11}. Following the method presented in \cite{dav13}, one can show that a given N$\ell$LO pseudopotential contributes to the central Landau parameters up to the multipolarity $\ell$, and to the tensor ones up to $(\ell-1)$. For instance, the standard Skyrme interaction is the N1LO pseudo-potential. To deal with future N3LO pseudo-potentials we shall truncate at $\ell_{max}=3$.

 \begin{table}[h!]
\begin{center}
\begin{tabular}{c|cccccc|ccc}
\hline
\hline
\multicolumn{10}{c}{Gogny D1MT \cite{ang11}}\\
\hline
&\multicolumn{6}{c|}{SNM} & \multicolumn{3}{c}{PNM}\\
\hline
$\ell$ & $F_{\ell}$ & $F'_{\ell}$ & $G_{\ell}$ & $G_{\ell}'$ & $H_{\ell}$ & $H_{\ell}'$&  $F^{(n)}_{\ell}$ & $G^{(n)}_{\ell}$ & $H^{(n)}_{\ell}$\\
0 & -0.310 & 0.724 & -0.037 & 0.731 & 0.306 & -0.102&  -0.560 &0.465  & 0.138 \\
1& -0.756 & 0.397 & -0.348 & 0.624& 0.587 & -0.196 &  -0.690 &  0.029&0.302\\
2& -0.293 & 0.615 & 0.471 & -0.237 & 0.557 & -0.185 &   0.283& 0.226 &0.329\\
3 &-0.055 & 0.130 & 0.108 & -0.060 &   -      &  -        &    0.120  & 0.077 & -     \\
\hline
\hline
\multicolumn{10}{c}{M3Y-P2 \cite{nak03}}\\
\hline
&\multicolumn{6}{c|}{SNM} & \multicolumn{3}{c}{PNM}\\
\hline
$\ell$ & $F_{\ell}$ & $F'_{\ell}$ & $G_{\ell}$ & $G_{\ell}'$ & $H_{\ell}$ & $H_{\ell}'$&  $F^{(n)}_{\ell}$ & $G^{(n)}_{\ell}$ & $H^{(n)}_{\ell}$ \\
\hline
0 &  -0.383 & 0.620 &  0.112 &  1.008  & 0.043& -0.015 & -0.564& 0.831&0.026\\
1& -1.040 & 0.632 & 0.273 & 0.201 &0.063 & -0.019      &    -0.362 &0.391&0.048\\
2& -0.433 & 0.243 & 0.161 & 0.040 &0.047 & -0.013       &   -0.174&0.223&0.044\\
3& -0.208 & 0.095 & 0.077 & -0.002&     -      &    -        & -0.106 &0.097& - \\
\hline
\hline                            
\end{tabular}
\caption{Landau parameters in SNM and PNM at density $\rho=0.16$ fm$^{-3}$ for the phenomenological interactions D1MT and M3Y-P2.}
\label{land-param-SNM}
\end{center}
\end{table}

In Table \ref{land-param-SNM} are collected the values of the Landau parameters, both for SNM and PNM, calculated from two types of phenomenological effective interactions. 
The first type  is based on the  effective density-dependent finite-range Gogny family interactions. In \cite{ang11}, the authors have provided two new interactions: D1ST  and D1MT. They correspond to the Gogny interaction D1S \cite{ber91}  and D1M~\cite{gor09} supplemented with a tensor term  obtained from the Argonne Av8' interaction \cite{wir02} in the following way. To take care of the short-range nucleon-nucleon (NN) correlations, the radial part of the tensor isospin term has been multiplied by a factor $\left(1-\exp (-b r_{12}^2)\right)$, and the parameter $b$ has been adjusted to fit the lowest $0^-$ states of some selected nuclei in a HF plus RPA description. 
An alternative definition for the finite range tensor term in the Gogny interaction has been presented in ~\cite{ang12}, based on the definition of Onishi and Negele~\cite{oni78}. The resulting interactions are labelled D1ST2a and D1ST2b.
In this case the tensor term is described with a single Gaussian of a fixed range with two strengths for each isospin channel, which 
have been adjusted to reproduce the experimental values of some shell gaps in calcium isotopes and the position of the first $0^{-}$ excited state in $^{16}$O.
Although the numerical value of the Landau parameters obtained from these four interactions are different, we have checked that they exhibit the same general features concerning the response of the system. We thus decided to focus on one interaction only. Our choice is D1MT due to better properties of the equation of state~\cite{cha08} especially in PNM, which is a case we studied in this paper.

The second type of interaction, labelled M3Y-P2 \cite{nak03}, is an effective density-dependent finite-range interaction which includes finite range tensor and spin-orbit components. It is a modification of the M3Y interaction, that was originally derived from a bare NN interaction by fitting Yukawa functions to a G-matrix \cite{ber77}.  All the parameters of this interaction have been obtained trough a complete fitting procedure including doubly magic nuclei and nuclear matter properties.

\begin{table}[h!]
\begin{center}
\begin{tabular}{c|cccc|cccc}
\hline
\hline
&\multicolumn{4}{c|}{CBF \cite{ben13}} & \multicolumn{4}{c}{CEFT \cite{hol13}}\\
\hline
$\ell$ & $F^{(n)}_{\ell}$ & $G^{(n)}_{\ell}$ & $\widetilde{H}^{(n)}_{\ell}$ & $H^{(n)}_{\ell}$ 
& $F^{(n)}_{\ell}$ & $G^{(n)}_{\ell}$ & $\widetilde{H}^{(n)}_{\ell}$ & $H^{(n)}_{\ell}$ \\
0 & -0.263 & 0.914 & 0.046 & 0.071 & 0.554 & 0.837 & 0.131 & 0.191  \\
1& -0.605 & 0.067 & 0.060 & 0.145 & -0.015 & 0.317 & 0.258  & 0.383 \\
2& -0.230 & 0.155 & 0.025 & 0.109 & -0.783 & 0.259 & -0.206 & 0.153 \\
\hline
\hline                            
\end{tabular}
\caption{Landau parameters in PNM at $\rho=0.16$ fm$^{-3}$ for the interaction of Benhar \emph{et al.}~\cite{ben13} and at  $k_F=1.7$ fm$^{-1}$ for the interaction of  Holt et al. \cite{hol13}. $\widetilde{H}^{(n)}_{\ell}$ are the originally calculated parameters, and $H^{(n)}_{\ell}$ are the values adapted to the definition used in this work. See text for details.}
\label{land-param-PNM}
\end{center}
\end{table}

For PNM we have also considered Landau parameters obtained from two recent microscopic calculations, given in Tab. \ref{land-param-PNM}.  Columns labelled CBF correspond to the results given in \cite{ben13}, where the formalism of the correlated basis function was applied using the Argonne V18 interaction \cite{wir95}. Columns labelled CEFT displays the parameters deduced in  the framework of chiral effective-field theory including two- and three-nucleon interactions \cite{hol13}. As mentioned above, these tensor parameters are based on a different definition than ours. Both sets of parameters are related by a recurrence relation \cite{ols04}, from which one can deduce the values required for the present formalism. Neglecting parameters with $\ell > 2$ we have obtained the values given in the last column of Tab. \ref{land-param-PNM} for each interaction. For consistency with our notation, the tildes have been exchanged with respect to Ref. \cite{ols04}. 

\subsection{Comparing response functions in the Landau approximation and in the RPA}

We have previously given explicit expressions for the response functions in the Landau approximation truncating the residual interaction to $\ell_{max}=1$. It is important to made the comparison with the RPA response functions calculated at the same truncation. 

\begin{figure}[h]
\begin{center}
\includegraphics[width=0.7\textwidth,angle=-90]{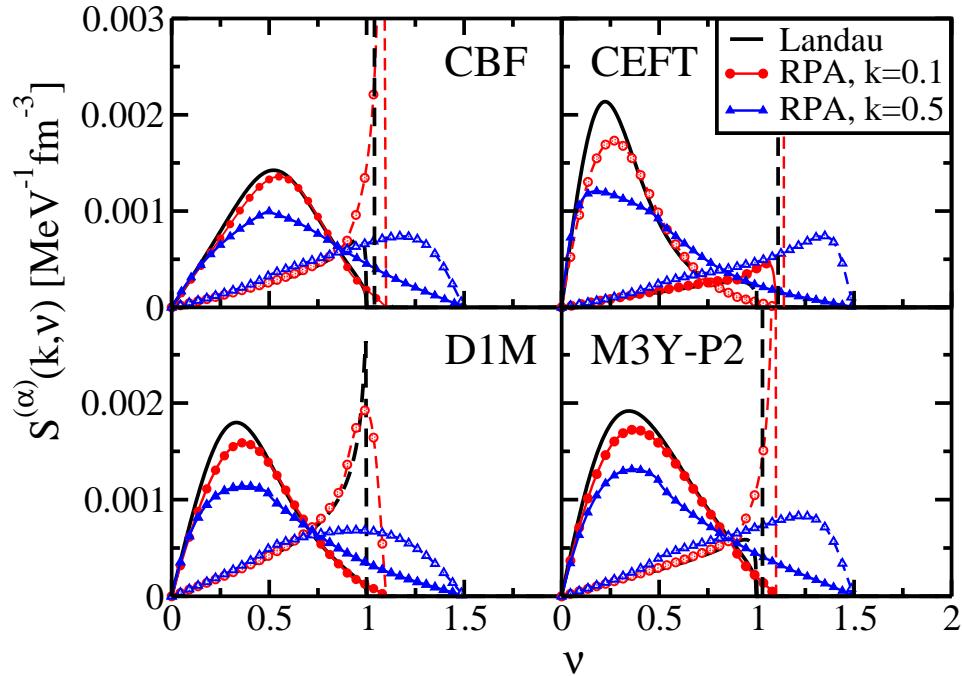}
\end{center}
\caption{(Color online) Pure neutron matter strength functions  calculated in the Landau approximation (Eq. \ref{LAN-01}) and in the RPA (Eq. \ref{RPA-01}) using the central $\ell=0$ and 1 parameters given in Tables \ref{land-param-SNM}-\ref{land-param-PNM}. Solid lines (with full symbols) and dashed lines (with open symbols) represent respectively channels $S=0$ and $S=1$. }
\label{fig:compPNM}
\end{figure}

Consider first the case of no tensor parameter. In Fig.~\ref{fig:compPNM} are plotted the PNM strength functions for both spin channels as a function of the dimensionless parameter $\nu$, calculated from the Landau approximation (\ref{LAN-01}) and from the RPA (\ref{RPA-01}). 
 Two values of the transferred momentum $q$ have been considered in the RPA case, $k=q/(2 k_F)= 0.1$ and 0.5. The results obtained with Landau parameters coming from CBF, CEFT, D1M, and M3Y-P2 interactions are displayed in different panels.

\begin{figure}[h]
\begin{center}
\includegraphics[width=0.6\textwidth,angle=-90]{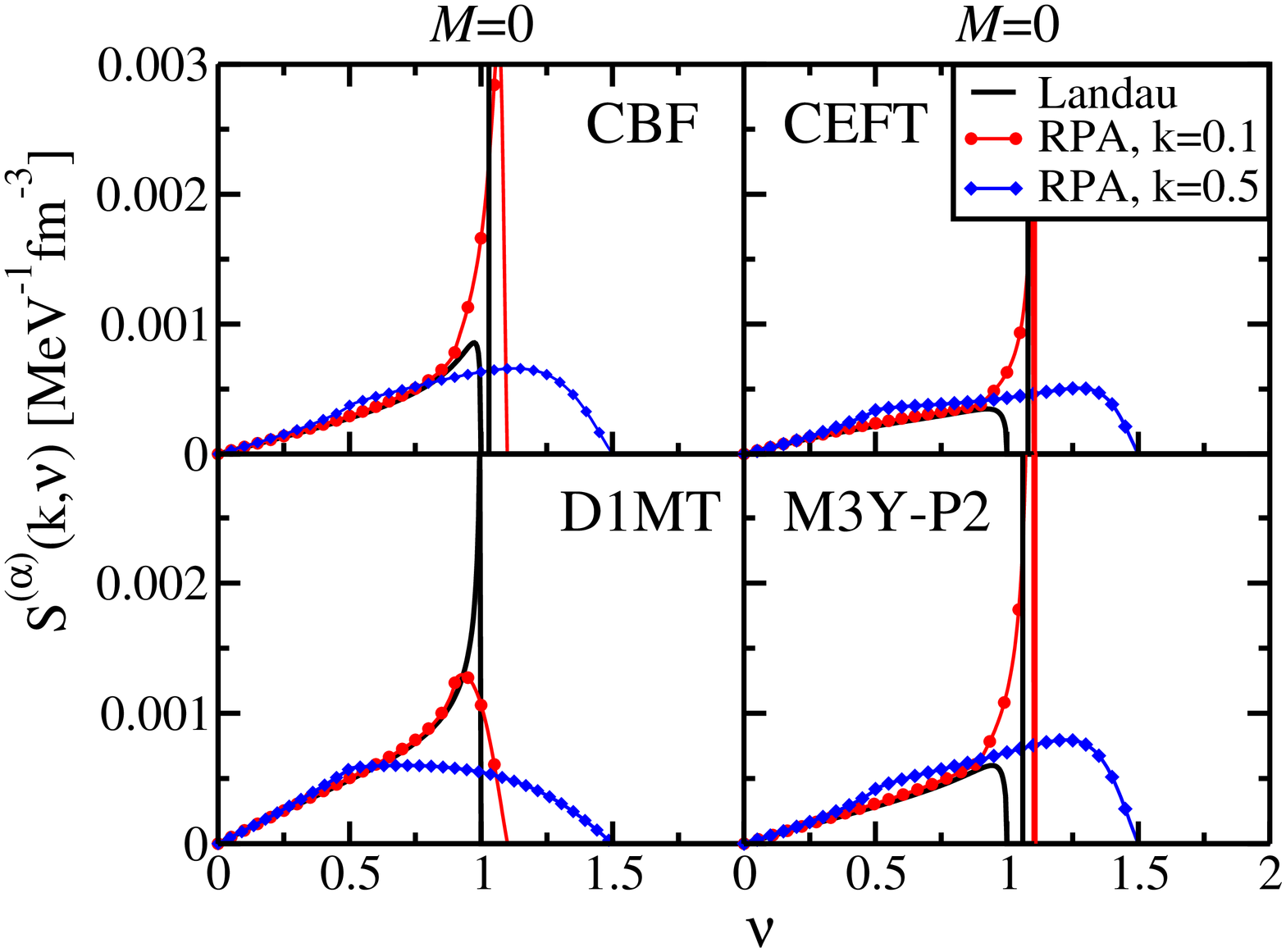}
\includegraphics[width=0.6\textwidth,angle=-90]{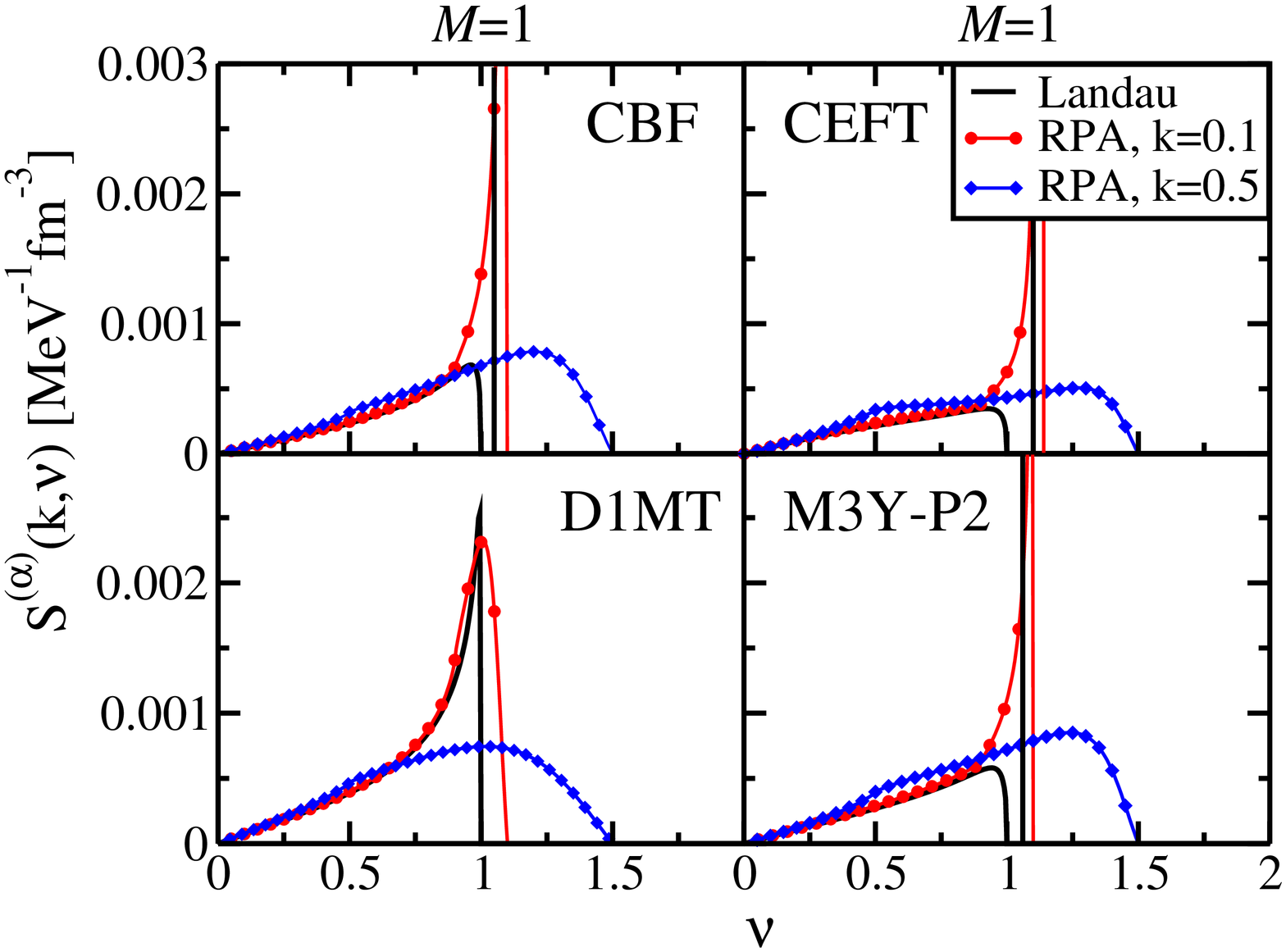}
\end{center}
\caption{(Color online) Pure neutron matter strength functions of $(S=1,M)$ channels calculated in the Landau approximation (Eq. \ref{LAN-01}) and in the RPA (Eq. \ref{RPA-01}) using the central $\ell=0$ and 1 parameters and tensor parameters $\ell=0$ given in Tables \ref{land-param-SNM} and \ref{land-param-PNM}. }
\label{fig:compPNM:2}
\end{figure}

It can be seen that the Landau response is nearly identical to the RPA one at $k=0.1$ and for values of $\nu$ below the peak. However, as soon as we consider larger values of the transferred momentum ($k=0.5$), there is an important redistribution of the RPA strength. In particular the zero-sound peak disappears, being absorbed into the continuum. The situation is not much different when the tensor parameter is included, as shown in Fig.  \ref{fig:compPNM:2}. Incidentally, one can observe that the tensor plays no significant role: Indeed, the responses in channels $(S=1,M=0)$ and $(S=1,M=1)$ look very similar. We cannot exclude that this is due to the small tensor strength of the considered interactions. In any case, this comparison establishes a quantitive limit to the validity of the Landau approximation for applications where the transferred momentum is not small, as for instance neutrino transport in dense matter. In the remaining sections, we shall present only RPA strength functions with a complete momentum dependence as in Eq. \ref{RPA-01} but including up to $\ell_{max}=3$.

\subsection{Symmetric Nuclear Matter RPA responses}

\begin{figure}[!h]
\begin{center}
\includegraphics[width=0.7\textwidth,angle=-90]{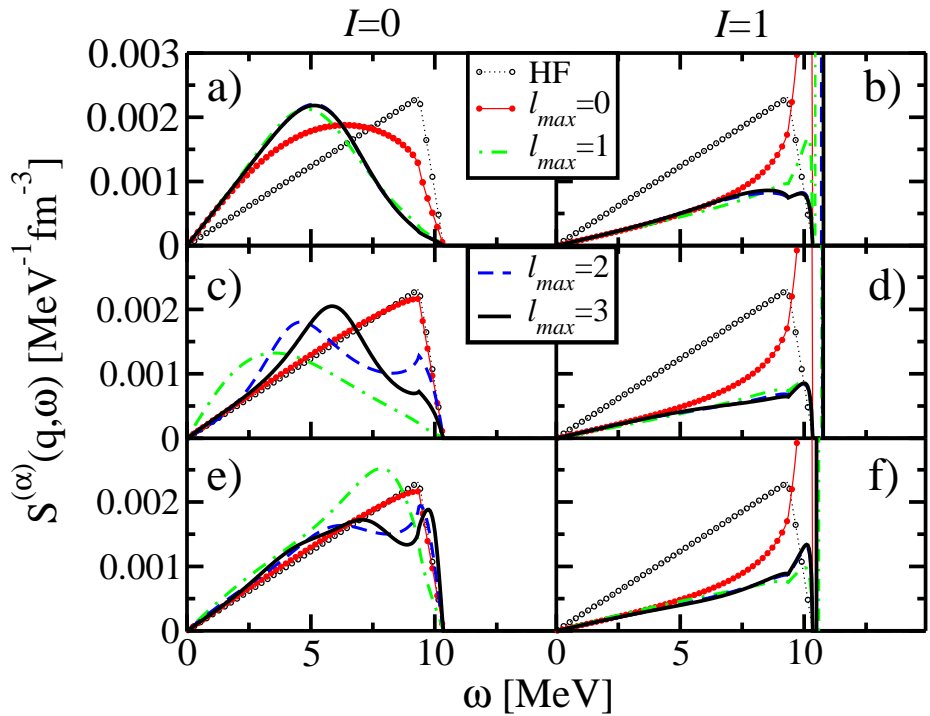}
\end{center}
\caption{(Color online) RPA strength functions in symmetric nuclear matter at density $\rho=0.16$ fm$^{-3}$ and $q=0.1 k_F$, calculated using D1MT parameters. Left and right panels displays the isospin channels $I=0$ and 1, respectively. The spin channels $S=0, M=0$ are plotted in the upper panels (a, b); $S=1, M=0$ in the middle panels (c,d); $S=1, M=1$ in the lower panels (e,f).}
\label{fig:cov:gognyD1MT}
\end{figure}

Let us discuss now the convergence of the RPA response functions for increasing values of $\ell_{max}$. For illustration we consider from now on the case of SNM in the six different spin-isospin channels. The strength functions for the D1MT interaction are shown in Fig. \ref{fig:cov:gognyD1MT} for a low value of transferred momentum  ($q/k_{F}=0.1$). To have a better insight of the effect of the ph interaction, we show on our figures the HF strength function  with the same value of the effective mass, namely $m^*/m=1+F_1/3$, with the D1MT value for the parameter $F_1$. Results for the $S=0$ channel are displayed in the upper panels (a, b), and one can see that a good convergence is obtained at $\ell_{max}=2$. 
Since there is no tensor contribution to the $S=0$ channel, we are just left with the central Landau parameters. This is one of the major differences compared to the results of \cite{dav09,pas12a,pas12b} obtained with the full Skyrme interaction, where due to the presence of a residual spin-orbit interaction, the tensor term contributes to both spin channels.
This coupling is suppressed in the Landau limit since there is no spin-orbit contribution to the ph interaction (\ref{landau-Vph}).
 The $S=1$ channel is more complex due to the explicit tensor contribution.
With the exception of the channel $S=1, M=0, I=0$ (panel c), the truncation at $\ell_{max}=2$ insures a good convergence. 

\begin{figure}[!h]
\begin{center}
\includegraphics[width=0.7\textwidth,angle=-90]{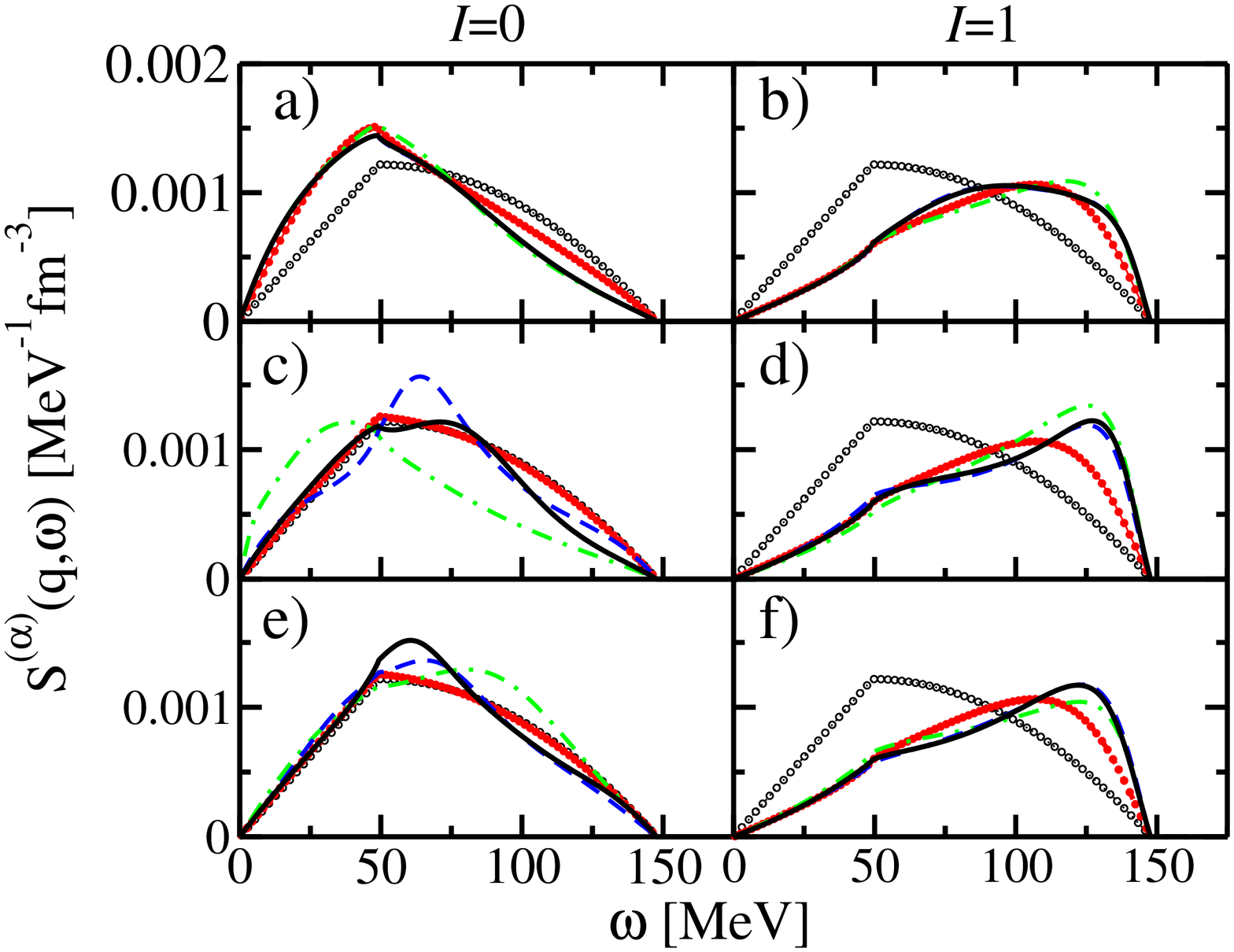}
\end{center}
\caption{(Color online) Same as Fig.\ref{fig:cov:gognyD1MT}, but for $q=k_F$. }
\label{fig:cov:gognyD1MT3}
\end{figure}

The same conclusions about the convergence hold for higher values of the transferred momentum.
By instance, Fig. \ref{fig:cov:gognyD1MT3}  shows the strength functions for $q= k_F$. Incidentally, one can notice that increasing the transferred momentum, the zero sound~\cite{nav99} observed in the $S=1$ channel disappears and is re-absorbed in the continuum part of the response function.

\begin{figure}[!h]
\begin{center}
\includegraphics[width=0.7\textwidth,angle=-90]{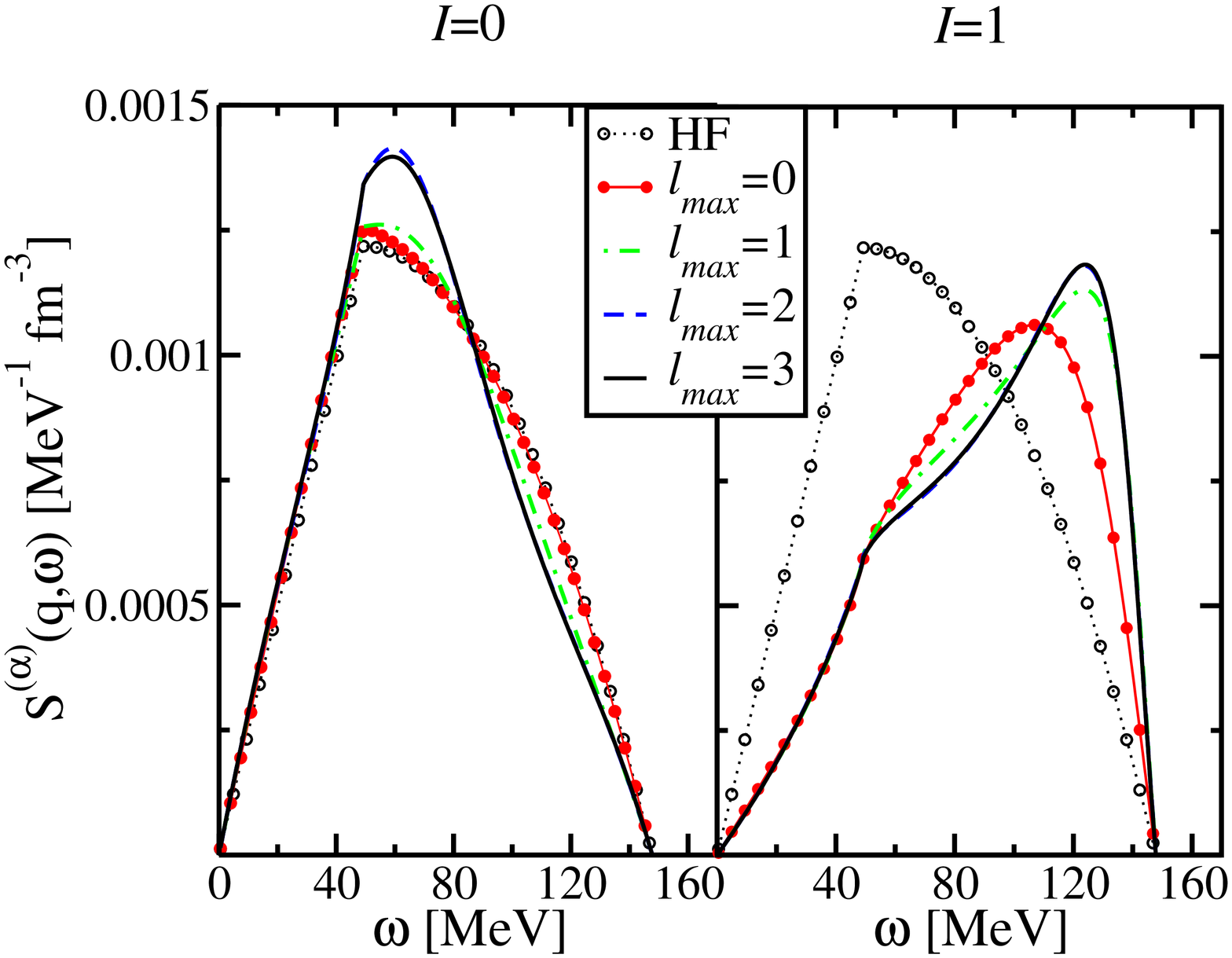} 
\end{center}
\caption{(Color online) Strength function in symmetric nuclear matter for the different isospin  channels at $S=1$ for the Gogny D1M~\cite{gor09} interaction at density $\rho=0.16$ fm$^{-3}$ and $q= k_F$ as in Fig.\ref{fig:cov:gognyD1MT3},  but with no tensor contribution. }
\label{fig:cov:gognyD1MT:h0}
\end{figure}

One should keep in mind that in D1MT the tensor interaction has not been derived from a complete re-fit of the parameters. One may wonder if the tensor terms could be responsible of the slow convergence  the $S=1, M=0, I=0$ channel. 
In Fig. \ref{fig:cov:gognyD1MT:h0} are plotted the $S=1$ strengths by putting all tensor terms to zero.
In this case the three spin-projections $M=0,\pm1$ are totally degenerate. Recall that this is a peculiarity of the Landau limit, otherwise, the two channels would be different because of the residual spin-orbit interaction.
We plainly observe that in this case we recover an excellent convergence at $\ell_{max}=2$, thus clearly demonstrating that the lack of convergence is related to the tensor terms.
Comparing the right panels ($I=1$) of Fig. \ref{fig:cov:gognyD1MT:h0} with the analogous of Fig. \ref{fig:cov:gognyD1MT} (d)-(f), we observe that the strength functions  with or without tensor are not very different. We can thus conclude that in this channel the tensor effects are small. However, this  conclusion cannot be reached from a simple inspection of the absolute values of the parameters $H_{l},H_{l}'$ given in Tab.\ref{land-param-SNM}: the effect is highly non linear and apart from general features, the behavior of the strength function can not be guessed directly from the numerical values of Landau parameters.

\begin{figure}[!h]
\begin{center}
\includegraphics[width=0.7\textwidth,angle=-90]{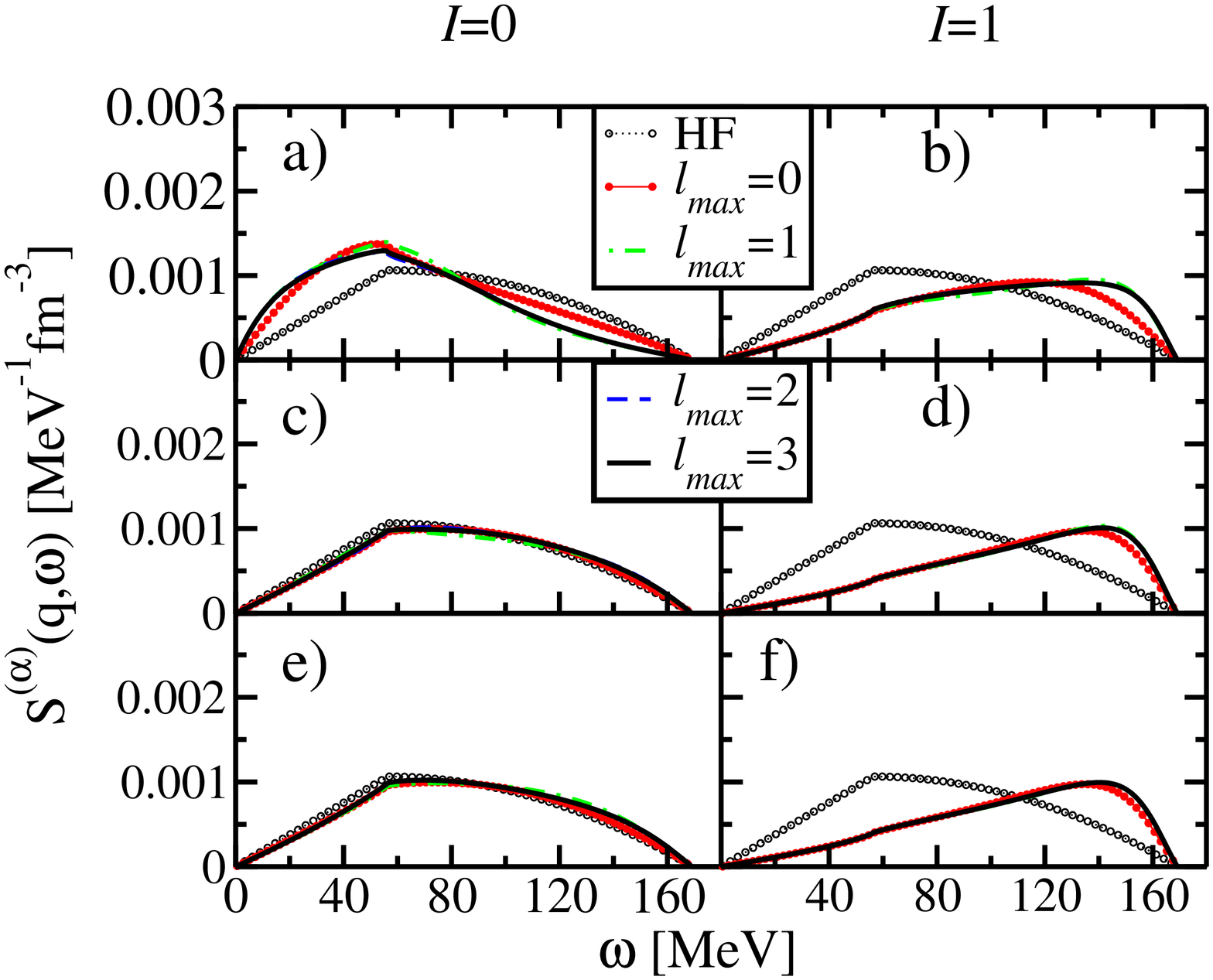}
\end{center}
\caption{(Color online) Strength function in symmetric nuclear matter for the different spin/isospin channels for the M3Y-P2  interaction at density $\rho=0.16$ fm$^{-3}$ and $q= k_F$.  In the different six panels we represent the channels $S=0,M=0,I=0,1$ panels (a)-(b), $S=1,M=0,I=0,1$ panels (c)-(d) and finally $S=1,M=1,I=0,1$ panels (e)-(f).}
\label{fig:cov:nakadaP2}
\end{figure}

In Fig. \ref{fig:cov:nakadaP2}, we present the convergence for the M3Y-P2 interaction~\cite{nak03}. 
With this interaction, the convergence as a function of $l$ seems to be on average better compared to D1MT.
Considering the Landau parameters of Tab. \ref{land-param-PNM}, we observe that there is one order of magnitude  for the values of $H_{l}$ in SNM between D1MT and M3Y-P2.
For M3Y-P2, tensor terms are sufficiently small in the isoscalar channel to basically give no contribution to the response function. This is clearly seen by comparing the two spin-projection $M=0$ and $M=1$,  and the HF response function, observing they are essentially identical.

\subsection{Pure Neutron Matter RPA responses}

As previously explained, we have written the expressions in such a way that they are immediately used for PNM calculations, by simply changing the values of the Landau parameters and the corresponding degeneracy of the system.

\begin{figure}[!h]
\begin{center}
\includegraphics[width=0.7\textwidth,angle=-90]{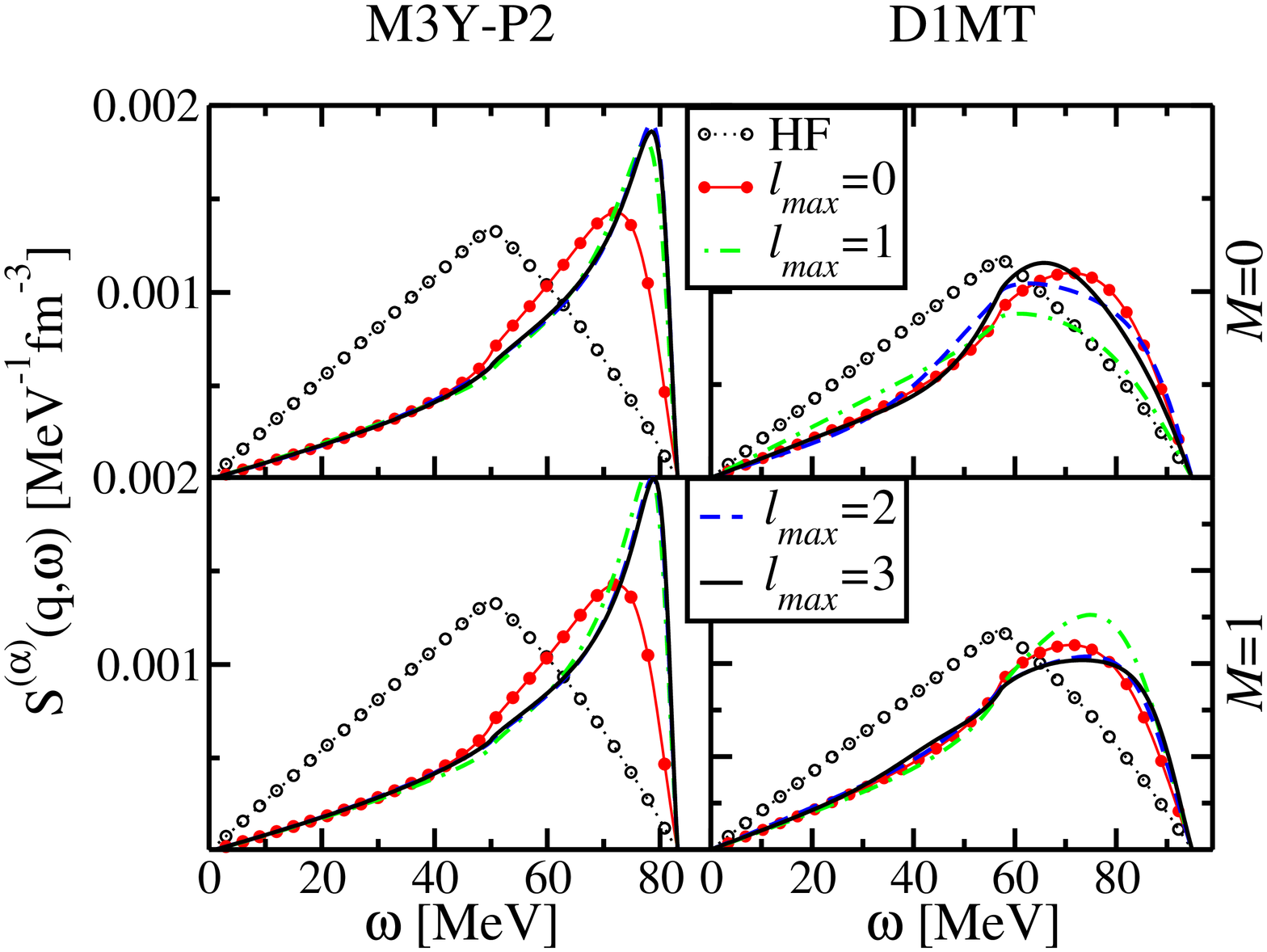}
\end{center}
\caption{(Color online) Strength function in pure neutron matter for $S=1$ channel for interactions M3Y-P2 and D1MT at density $\rho=0.16$ fm$^{-3}$ and $q=0.5 k_F$.}
\label{fig:cov:gognyD1MT:pnm}
\end{figure}

In Fig. \ref{fig:cov:gognyD1MT:pnm}, we present strength functions in the $S=1$ channel at density $\rho=0.16$ fm$^{-3}$ and for $q=0.5 k_F$. We do not consider the $S=0$, because the tensor plays no role. The interactions are M3Y-P2 and D1MT. We observe that in the latter case the strength functions are mainly dominated by central terms, and including or not the tensor is not important since the $M=0$ and $M=1$ channels are quite similar. We clearly see that in both channels, convergence is achieved for $\ell_{max}=2$.
We notice that the strength functions obtained with M3Y-P2 and D1MT are quite similar in these channels. This feature can be understood from Tab. \ref{land-param-SNM}. Even if it is difficult to draw some conclusions from the numerical values, one can see that the parameters $G^{(n)}_{\ell}$ are essentially the same for the two interactions, and since the tensor terms do not give a strong contribution, the resulting response functions are essentially the same.

\begin{figure}[!h]
\begin{center}
\includegraphics[width=0.7\textwidth,angle=-90]{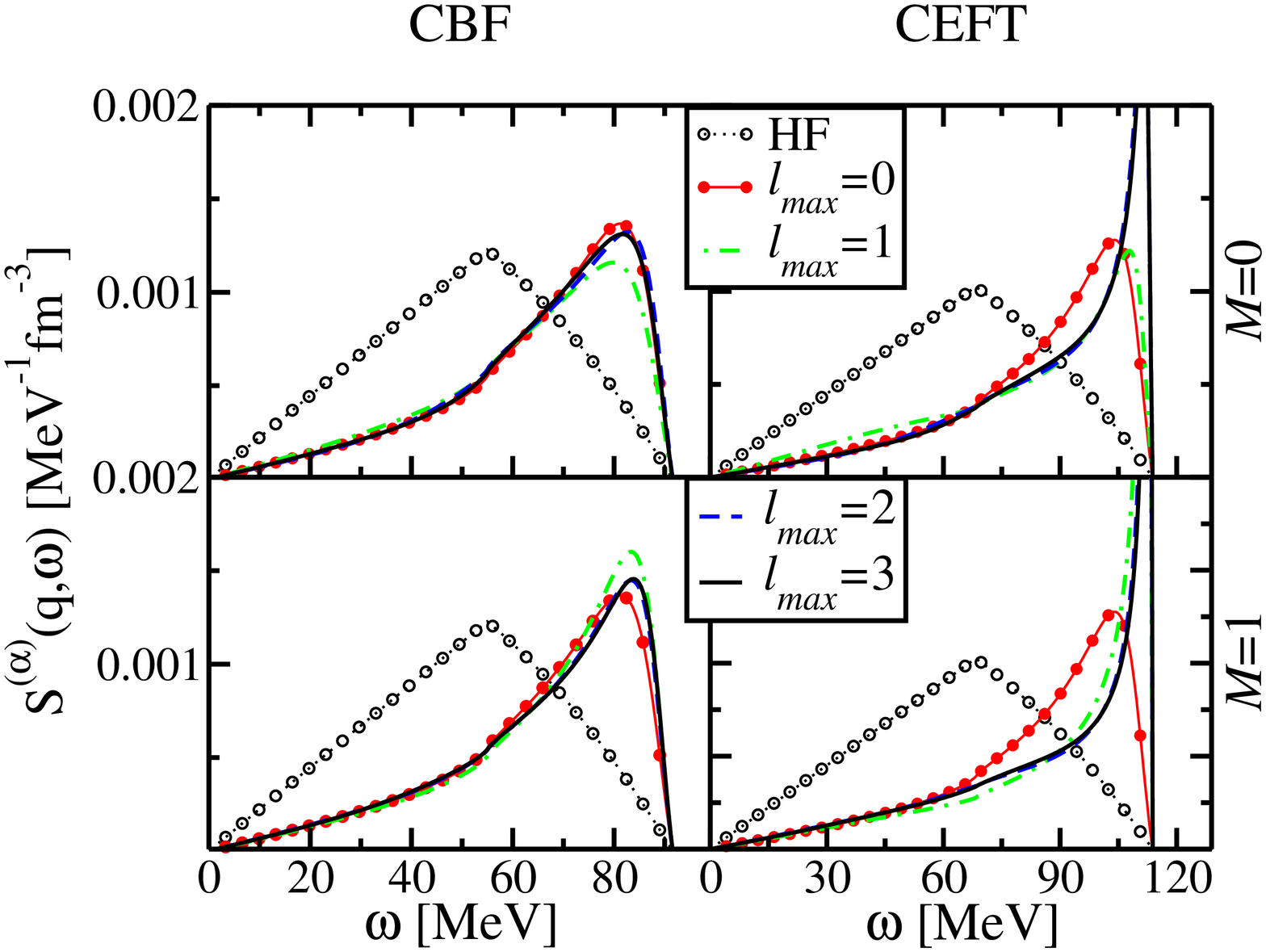}
\end{center}
\caption{(Color online) Strength function in pure neutron matter for $S=1$ channel at $q=0.5 k_F$ for interactions CBF \cite{ben13} at density $\rho=0.16$ fm$^{-3}$ and CEFT \cite{hol13}  at $k_F=1.7$ fm$^{-1}$.}
\label{fig:cov:cbf}
\end{figure}

Finally, in Fig. \ref{fig:cov:cbf} the strength functions obtained from  CBF and CEFT interactions
are also depicted using the tensor 
parameters given in the last columns of Tab. \ref{land-param-PNM}. We remind that we are using approximated values for tensor parameters, as they have been obtained truncating a recurrence relation. These interactions produce very different responses in 
 both $(S=1, M)$ channels. In particular, CEFT predicts a narrow resonance, whereas CBF displays a broad structure. 
 However in both cases tensor effects are not very significant, since the response functions are quite similar for $M=0$ and $M=1$. 
 Finally, the convergence is attained for $\ell_{max}=2$ like the previous studied cases.

\section{Conclusions}\label{conclusion}

We have derived RPA response functions for the case of a residual interaction described in terms of its Landau parameters, including tensor contributions. The response functions in the Landau approximation have been immediately obtained from the RPA ones by taking the long-wavelength limit keeping $\nu$ finite. In this way, analytical expressions have been given for the Landau responses up to $\ell_{max}= 1$ (i.e. the first two central and the first tensor parameters). For other cases, and in general for the RPA responses, it is useless trying to get analytical expressions and one has to solve numerically an algebraic system of linear equations.

Both responses coincide as expected for low values of the transferred momentum (typically $q \leq 0.2 k_F$) and low values of $\nu$. However neat differences are visible when $q$ increases. The RPA strength displays an important reshaping, in particular the absorption of the zero sound peak into the continuum. This conclusion is particularly interesting if one considers quantities such as the neutrino mean free path in dense matter, since it requires the knowledge of the response function at transferred momentum values which are not small.

We have also studied the convergence as the number of Landau parameters is increased. Both in SNM and PNM, we have seen that a satisfactory convergence is reached for $\ell_{max}=2$ for the considered interactions. The exception is the $(S,M,I)=(1,0,0)$ channel and D1MT interaction, which has been explained by the relative importance of tensor terms in that particular case.

The response functions presented in this paper have an universal character since they can be related to any NN-potential through their Landau parameters. Therefore, our results could be useful for new generation of low energy effective interactions or for higher-order Skyrme-type potentials (N3LO). Concerning the latter case, this work suggests the possibility to use Landau parameters to constrain higher order terms, and in particular tensor terms. The latter are very difficult to constrain using only ground state properties, thus the use of Landau response functions could be of help to define the correct interval of variation of such terms, thus avoiding instability regions~\cite{mar02,nav13}. However, the latter conditions are necessary but not sufficient to have a stable Skyrme pseudo-potential: as recently shown in ~\cite{les07,fra12}, poorly-constrained coupling constants related to gradient terms can produce finite-size instabilities.
These instabilities can be detected using the linear response formalism~\cite{pas12c,pas13c,hel14}, with no extra computational cost and directly included into the optimization procedure~\cite{kor10}.

\section*{Acknowledgments}
This work was supported by NESQ project (ANR-BLANC 0407, France) and Mineco (Spain), grant FIS2011-28617-C02-02. The authors thank  M. Ericson and S. Goriely  for useful comments. 

\appendix

\section{$\alpha_{i=0,8}(k,\nu)$ functions} \label{app:alpha}

In this section we give the explicit expression of the functions $\alpha_{i=0,8}$ used in the text.
Notice that they are expressed in terms of the dimensionless quantities $k=q/(2k_{F})$ and $\nu=m^{*}\omega/(2 k_{F}^{2} k)$.
For indices $i=0-2$ we have:
\begin{eqnarray}\label{eq:alpha}
\fl  {\rm Im} \, \alpha_{0}(k,\nu) &=& -\frac{\pi}{2} \left[(1-(\nu-k)^{2}) \, \Theta_{-} -(1-(\nu+k)^{2}) \, \Theta_{+} \right] \\
\fl   {\rm Im} \, \alpha_{1}(k,\nu) &=& -\pi (\nu-k)\left[(1-|\nu-k|) \, \Theta_{-} -(\sqrt{1-4k\nu}-|\nu-k|) \, \Theta_{+} \right] \\
\fl   {\rm Im} \, \alpha_{2}(k,\nu) &=& \pi (\nu-k)^{2}\left[\log |\nu-k| \, \Theta_{-} +\log \frac{\sqrt{1-4k\nu}}{|\nu-k|} \, \Theta_{+} \right] 
\end{eqnarray}
For $i \ge 3$:
\begin{eqnarray}
\fl   {\rm Im} \, \alpha_{i}(k,\nu) &=& \frac{\pi}{i-2}(\nu-k)^i \left[\left(1-\frac{1}{|\nu-k|^{i-2}}\right) \, \Theta_{-} \right. \\
&& \left. \hspace{3cm} + \left(\frac{1}{|\nu-k|^{i-2}}-\frac{1}{(1-4k\nu)^{(i-2)/2}}\right) \, \Theta_{+}  \right] \nonumber
\end{eqnarray}
where we have defined $\Theta_{\pm}=\Theta(1-(\nu\pm k)^{2})$, $\Theta$ being the standard step function.
The real parts are obtained by mean of a standard dispersion relation:
\begin{eqnarray}
 {\rm Re} \, \alpha_{i}(\nu,k)=-\frac{1}{\pi}\int_{-\infty}^{+\infty} d \nu'  \, \, \frac{ {\rm Im} \, \alpha_{i}(k,\nu') }{\nu-\nu'}
\end{eqnarray}

\begin{figure}[h!]
\begin{center}
\includegraphics[angle=-90,width=0.49\textwidth]{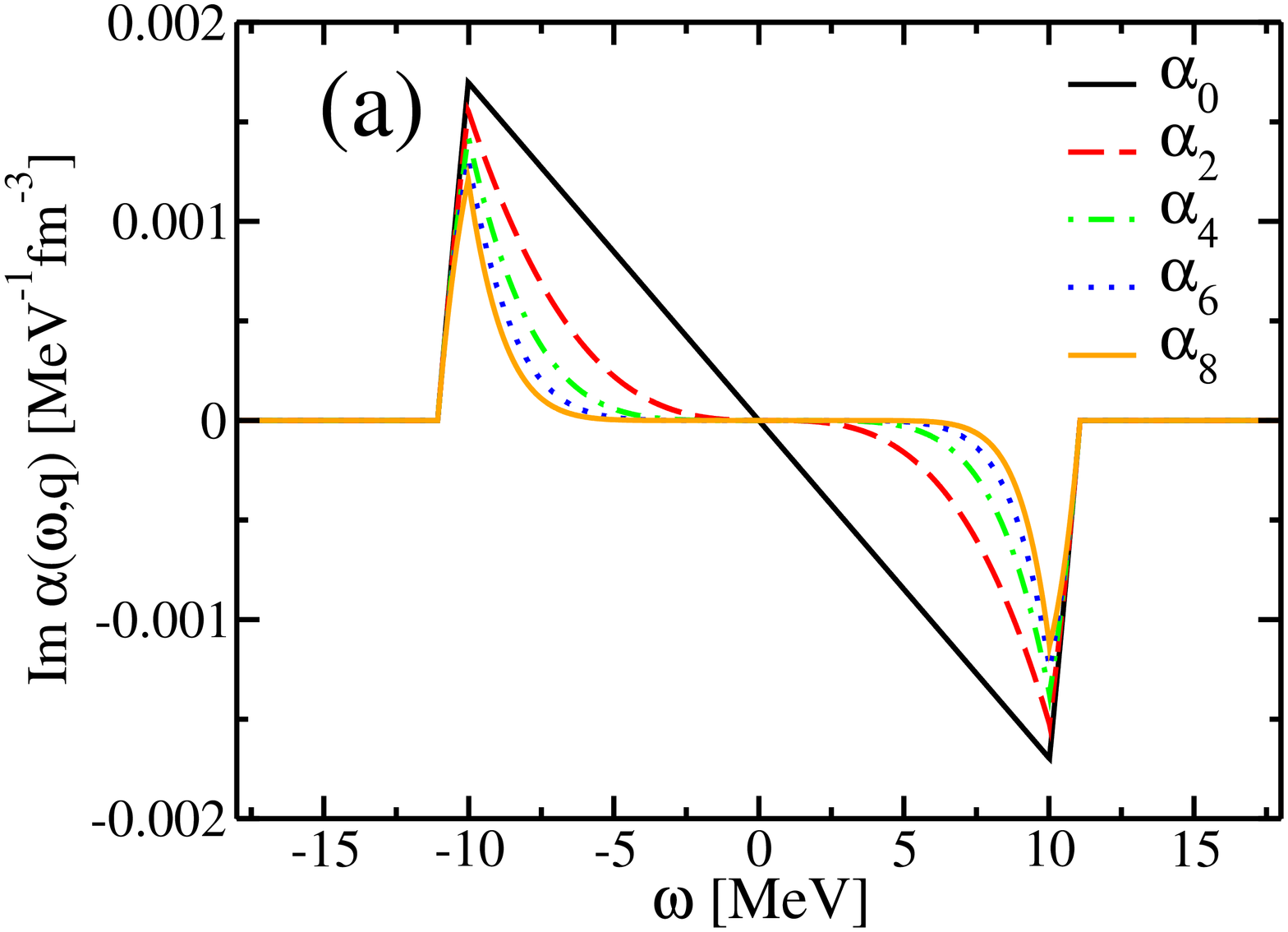}
\includegraphics[angle=-90,width=0.49\textwidth]{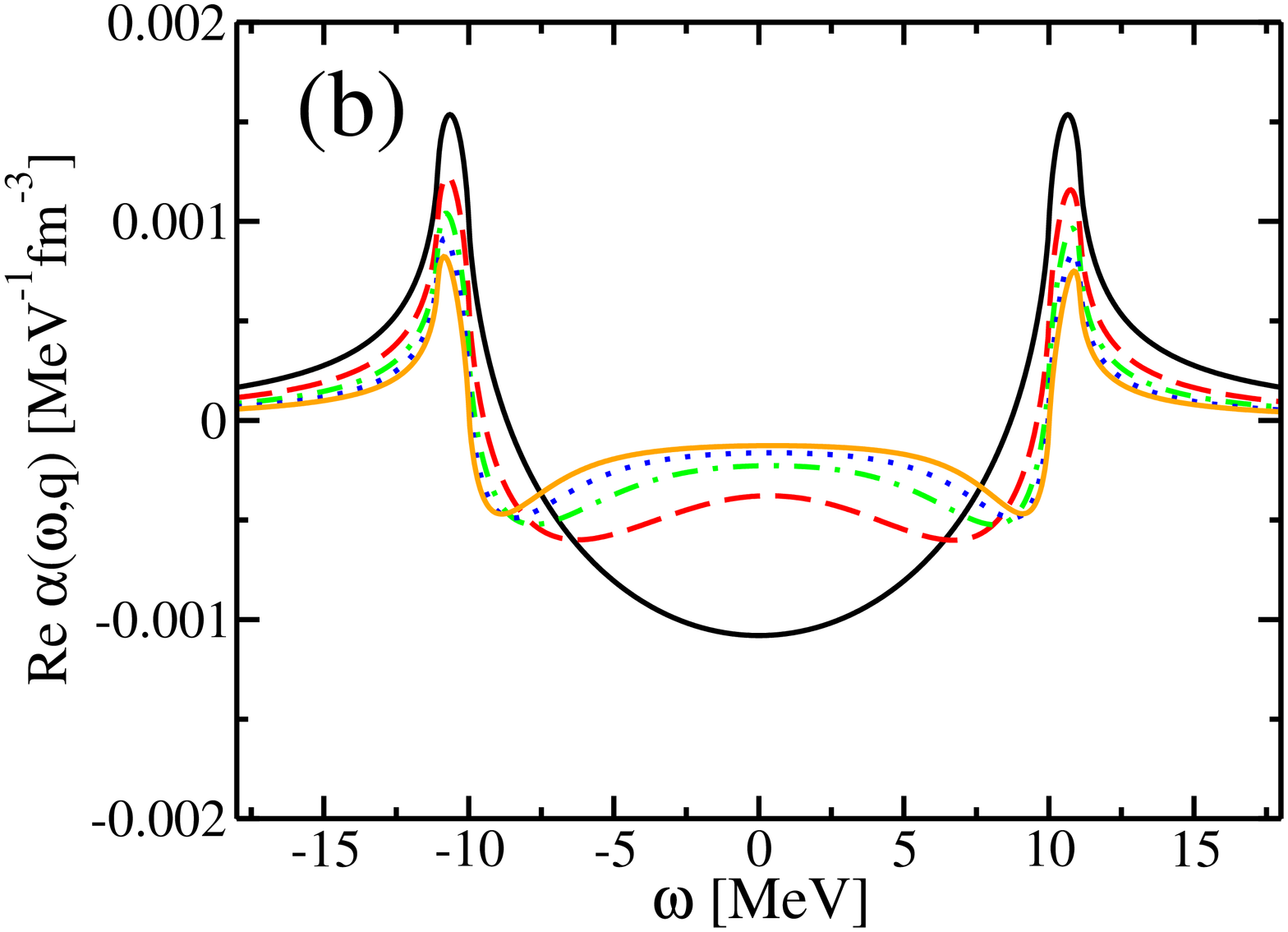}
\end{center}
\caption{(Color online) We represent the even $\alpha_{i=0,2,4,6,8}(k,\nu)$ functions On panel (a) (respectively panel (b)), imaginary parts (respectively real parts) are depicted. The calculations have been done for an effective mass of $m^{*}/m=0.7$ and $\rho=0.16$ fm$^{-3}$ in SNM using $q/k_{F}=0.1$.}
\label{fig:alphaE}
\end{figure}

\begin{figure}[h!]
\begin{center}
\includegraphics[angle=-90,width=0.49\textwidth]{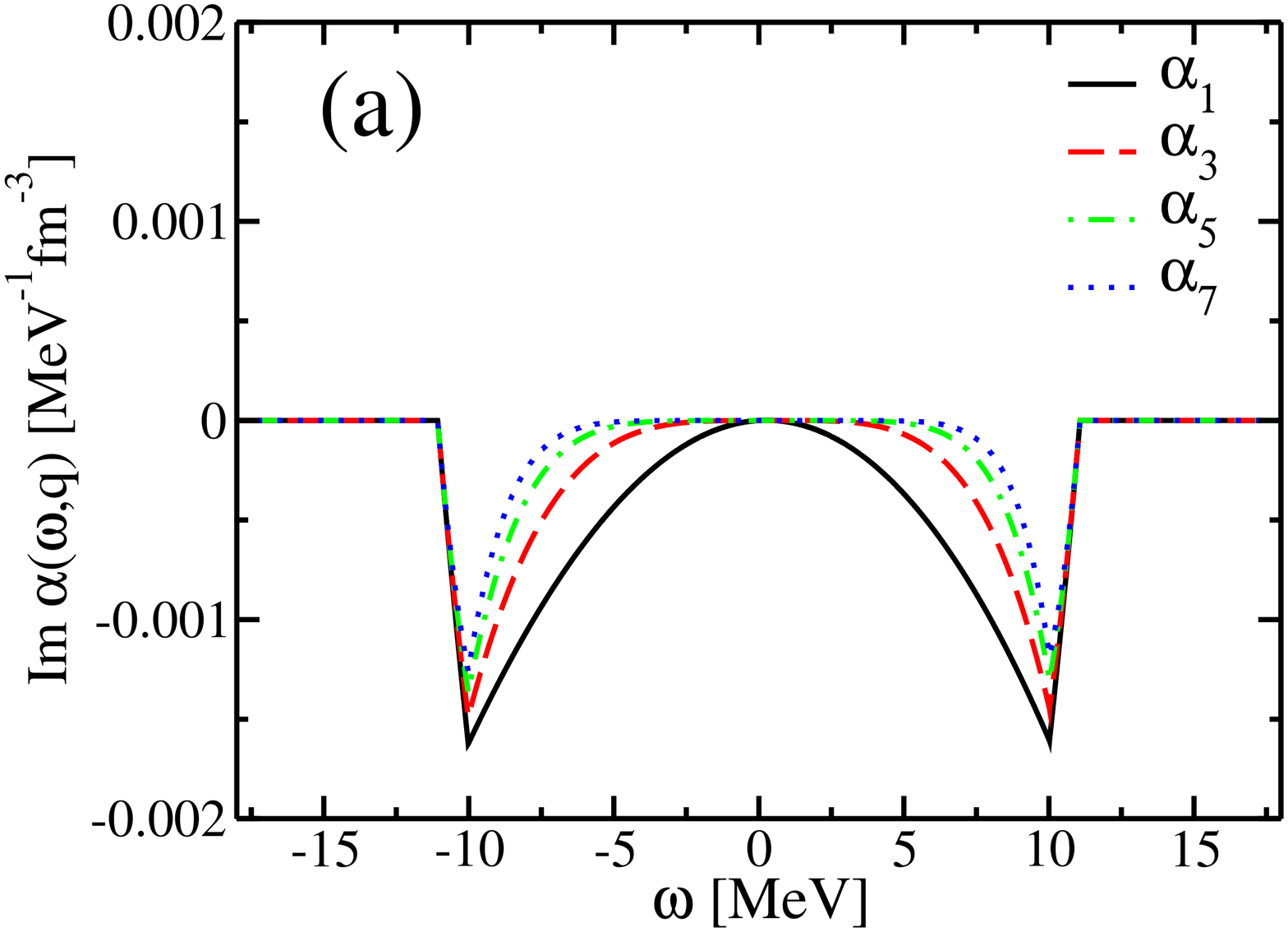}
\includegraphics[angle=-90,width=0.49\textwidth]{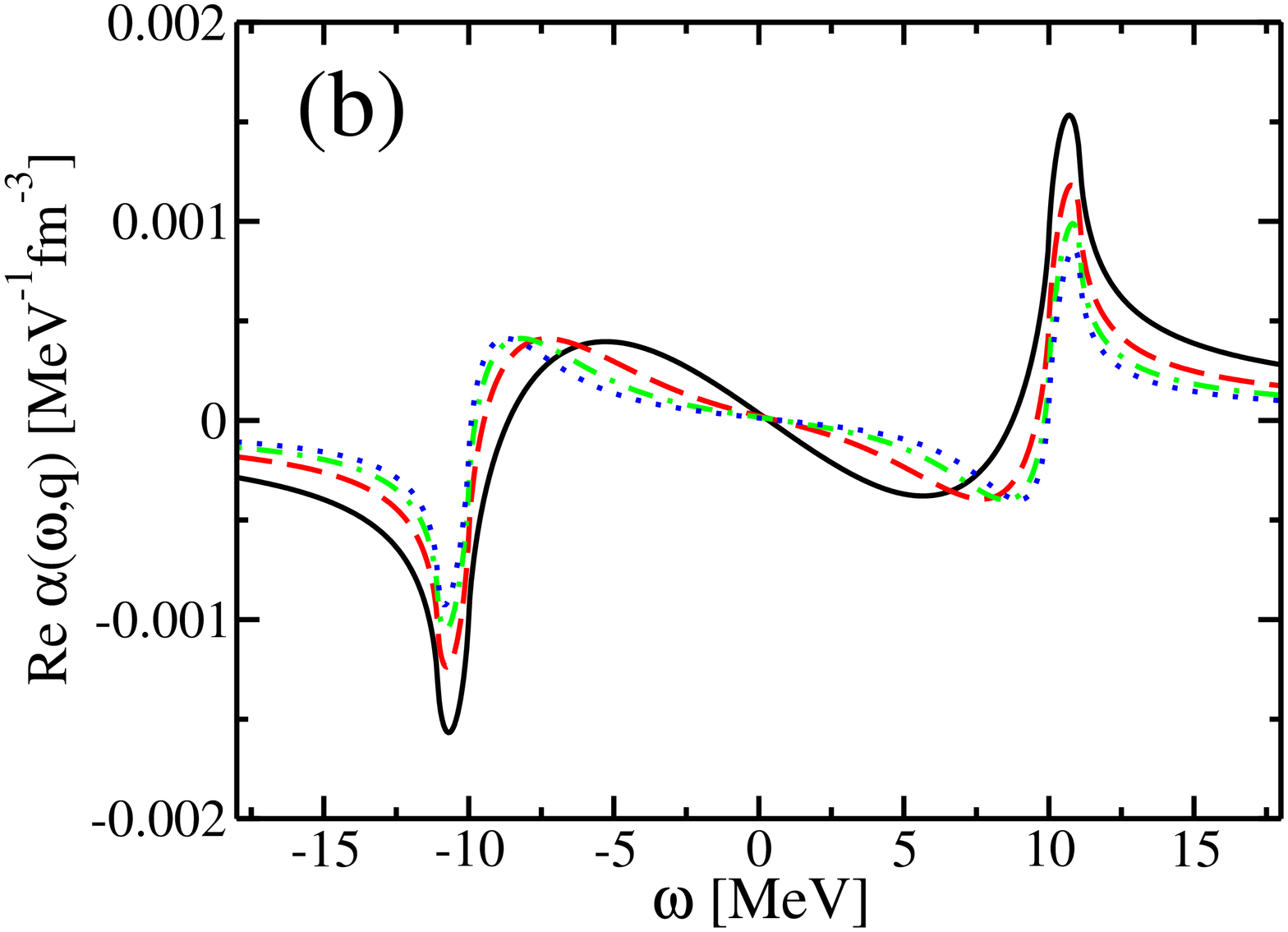}
\end{center}
\caption{(Color online) Same as Fig.\ref{fig:alphaE}, but for odd values of $i=1,3,5,7$.}
\label{fig:alphaO}
\end{figure}

In Figs.~\ref{fig:alphaE}-\ref{fig:alphaO}, we give an example of the functions $\alpha_{i}(k,\nu)$ calculated using an effective mass of  $m^{*}/m=0.7$  at $\rho=0.16$ fm$^{-3}$.

\end{document}